\documentclass[pra,aps,showpacs,preprint]{revtex4}
\usepackage{dcolumn}

\begin{document}

\input epsf

\preprint{FNAL-Pub-00/206-A \ \ IAS-SNS-AST-0035}

\topmargin 0.0in
\def\simlt{\stackrel{<}{{}_\sim}}
\def\simgt{\stackrel{>}{{}_\sim}}
\def\sfrac#1#2{{\textstyle{\sst{#1}\over\sst{#2}}}}
\newcommand{\Frac}[2]{\frac{{\displaystyle #1}}{{\displaystyle #2}}}
\renewcommand{\arraystretch}{2}

\def\cm{{\rm cm}}
\def\s{{\rm s}}
\def\GeV{{\rm GeV}}
\def\km{{\rm km}}
\def\g{{\rm g}}
\def\K{{\rm K}}
\def\yr{{\rm year}}

\newcommand\simp{{\sc simpzilla}}
\newcommand\simps{{\sc simpzillas}}

\newcommand\nt{$\nu_{\tau}$}

\title{High Energy Neutrinos From \\ Superheavy Dark Matter Annihilation}

\vspace*{-24pt}

\author{Ivone F.\ M.\ Albuquerque}\thanks{Electronic mail: ifreire@fnal.gov}
\affiliation{Space Science Laboratory and Astronomy Department, 
        University of California, Berkeley, California \ 94720 }

\author{Lam Hui}\thanks{Electronic mail: lhui@ias.edu}
\affiliation{Institute for Advanced Study, School of Natural Sciences, 
	Einstein Drive, Princeton, 
	New Jersey \  08540,\\
	and Department of Physics,
	Columbia University, 538 West 120th Street, New York, New York 10027}

\author{Edward W.\ Kolb}\thanks{Electronic mail: rocky@fnal.gov}
\affiliation{NASA/Fermilab Astrophysics Center, Fermi
        National Accelerator Laboratory, Batavia, Illinois \
        60510-0500,\\ and Department of Astronomy and Astrophysics,
        Enrico Fermi Institute, \\ The University of Chicago, Chicago,
        Illinois \ 60637-1433}

\date{August 2000}

\begin{abstract}
Superheavy ($M>10^{10}$ GeV) particles produced during inflation may
be the dark matter, independent of their interaction strength.
Strongly interacting superheavy particles will be captured by the sun,
and their annihilation in the center of the sun will produce a flux of
energetic neutrinos that should be detectable by neutrino
telescopes. Depending on the particle mass, event rates in a
cubic-kilometer detector range from several per hour to several per
year.  The signature of the process is a predominance of tau
neutrinos, with a relatively flat energy spectrum of events ranging
from 50 GeV to many TeV, and with the mean energy of detected tau
neutrinos about 3 TeV.
\end{abstract}


\maketitle

\section{Introduction}

It is usually assumed that the interactions of dark matter and
ordinary matter are weak, at least as small as ordinary weak
interactions.  It is also often assumed that the dark matter is a
thermal relic of the big bang.  If it is a thermal relic particle,
then its mass should be less than 340 TeV, the unitarity limit
\cite{gk}.  These considerations lead to the popular picture for relic
dark matter of an an electrically neutral particle, without strong
interactions, and with mass less than a couple of hundred TeV.  In
this paper we explore a path less traveled, and assume that the dark
matter is a nonthermal relic, it interacts strongly with normal
matter, and it is very massive.  We show that the clean signature of
this possibility is a detectable flux of energetic neutrinos from
the sun

It has long been appreciated that if the dark matter is massive, say
larger than a few TeV, it will behave effectively as dissipationless
dark matter regardless of whether it has strong or electromagnetic
interactions \cite{al,stark}.  However, since the upper limit to the
mass of a thermal relic is a few hundred TeV, the window for very
massive dark matter particles was thought to be not very wide.

The recent development of scenarios for nonthermal production of dark
matter has opened the window to the possibility that the dark matter
might be supermassive, independent of its interaction strength
\cite{ckr,kt,hs99}.  Of the many possibilities for producing supermassive 
dark matter, perhaps gravitational production is the most general
\cite{ckr,kt}.  In this scenario, dark matter is produced by vacuum
quantum fluctuations toward the end of inflation.  The resulting
particle density is independent of the interaction strength of the
particle, which leads to the possibility that the dark matter may be
electrically charged, strongly interacting, weakly interacting, or may
have only gravitational interactions with normal matter.

A particularly promising mass range for gravitational production of
dark matter is the mass scale of the inflaton, about $10^{12}$ GeV in
chaotic inflation models.  If the inflaton mass heralds a new mass
scale, then it would be reasonable to imagine that there are other
particles of similar mass.  Furthermore, gravitational production of
particles with a mass comparable to the inflaton mass naturally leaves
behind a cosmologically interesting density of dark matter today
\cite{ckr}.  The particle content may include exotic quarks or other
strongly interacting particles in the spectrum of new particles
\cite{simps}.

In this paper we will consider the case that the dark matter is
strongly interacting and supermassive, a \simp.  Although our
calculations will not be sensitive to whether the \simp\ is
electrically charged, there are arguments that suggest that the \simp\
must be neutral \cite{neutronstars,esteban}.  The possibility that the
dark matter may be very massive and strongly interacting was recently
discussed by Faraggi, Olive, and Pospelov \cite{boykeith}.  In that
paper they have a nice discussion of the particle physics motivations
for the existence of a massive, stable, strongly interacting particle,
and they point out that the sun and Earth may be the source of
high-energy neutrinos from the annihilation of the particles.

In the next section we will calculate the trapping rate and
annihilation rate of \simps\ in the sun (and Earth) as a function of
the \simp\ mass and interaction cross section.  In Section III we will
discuss the emergent spectrum of neutrinos from \simp\ annihilation in
the center of the sun. In Section IV we will calculate the event rate in
cubic-kilometer underwater or underice neutrino detectors.  Finally,
the last section contains our conclusions.

A preview of our main conclusion is that we expect a detectable
high-energy solar neutrino flux for much of the parameter range of
interest.

\section{Capture and Annihilation Rate}

Before launching into the details of the capture-rate calculation it
is useful to make some extremely crude estimates.  The first estimate
is for the number of \simps\ that hit the sun.

Assuming the \simps\ comprise the local dark matter density of $0.3\
\GeV\ \cm^{-3}$, the local number density of \simps\ is
$3\times10^{-13}\ M_{12}^{-1}\cm^{-3}$, where $M_{12}$ is the \simp\
mass in units of $10^{12}\GeV$. Assuming a typical velocity of $240\
\km\ \s^{-1}$ for the \simps, the local flux is approximately
$6\times10^{-7}M_{12}^{-1}\cm^{-2}\s^{-1}$.  The surface area of the
sun is about $6\times10^{22}\cm^2$, and the product of the surface
area and the flux, $4\times 10^{16}M_{12}^{-1}\s^{-1}$, is a crude
estimate of the rate of \simps\ hitting the sun.  

Now consider trapping of \simps\ in the sun.  Assuming the \simp\
impacts the sun with the solar escape velocity of $600\ \km\ \s^{-1}$,
the kinetic energy of the \simp\ is about $10^6M_{12}\GeV$.  Suppose
the \simp\ scatters with nucleons with a cross section of
$\sigma=10^{-24}\sigma_{-24}\ \cm^2$ \cite{crosssection}, and in every
collision suffers an energy loss of $m_{\rm proton} v^2/2 =
2\times10^{-6}\GeV$. Through the center of the sun is about
$2\times10^{36}\cm^{-2}$ of material, so there will be about
$2\times10^{12}\sigma_{-24}$ collisions with a total energy loss of
$4\times10^6\sigma_{-24}\GeV$.  Since the initial kinetic energy is
about $10^6M_{12}\GeV$, for masses much less than
$10^{12}\sigma^{-1}_{-24}\GeV$, \simps\ with average velocity hitting
the sun will be trapped, but if the \simp\ mass is much larger than
that, only the low-velocity tail of the phase-space distribution will
be captured.

These considerations determine the gross behavior of the dependence of
the \simp\ trapping rate on the \simp\ mass and its interaction cross
section.  For a mass of about $10^{12}\GeV$, about $10^{17}$ \simps\
hit the sun per second. For an interaction cross section of about
$10^{-24}\cm^2$, most of those \simps\ are captured.

Now we turn to the details of the capture calculation.  The capture of
dark matter particles in the sun and Earth has been studied in detail
\cite{gould87,pressnspergel}; here we adapt the considerations to the
capture of \simps\ by the sun.  The capture rate of dark matter by the
sun is given in terms of an integral over $f(u)$, the phase-space
density of dark matter in the solar neighborhood normalized such that
$\int_0^\infty f(u) du$ gives the number density of particles at $R$,
some sufficiently large radius where the gravitational pull of the sun
is negligible.  The capture rate is given by
\begin{equation}
\Gamma_C = 4 \pi R^2 \int \left[ \frac{1}{4} f(u) \, u \, du 
\ d\sin^2\theta \right] P(\theta, u) ,
\label{C}
\end{equation}
where $\theta$ is the angle between the velocity of the particle and
the normal to the surface at $R$, and hence the expression inside the
square bracket is the contribution to the inward flux at radius $R$
from particles at velocity $u$ and angle $\theta$.  Finally,
$P(\theta, u)$ gives the probability that particles with the
given velocity and direction at some large radius $R$ will be captured
by the sun.

Writing the angular momentum per unit mass as $J = Ru\sin\theta$, and
defining $w_\odot^2 = u^2 + v_\odot^2$ where $v_\odot$ is the escape
velocity at the surface of the sun, we can rewrite the above as
\begin{equation}
\Gamma_C = \pi \int_0^\infty f(u)\, u \, du  \int_0^{R_\odot w_\odot}
\frac{dJ^2}{u^2} P(J, u)  ,
\label{C2}
\end{equation}
where $R_\odot$ is the solar radius, and $P(J,u)$ now gives the
capture probability for a particle with angular momentum $J$ and a
velocity at infinity of $u$.

The regime in which dark matter capture is often considered is that of
weakly interacting particles, where the optical depth of the sun is
much less than unity, {\em i.e.,} $n_N \sigma R_\odot \ll 1$ where
$n_N$ is the number density of nuclei in the sun, and $\sigma$ is the
dark-matter---nucleon cross section. In this case $P(J,u)$ is given by
\cite{gould87}:
\begin{equation}
\label{P}
P(J,u) =  \left(\int_0^{R_\odot w_\odot} \frac{dJ^2}{u^2}\right)^{-1}
\int_0^{R_\odot} dr \int_0^{rw} \frac{dJ^2}{u^2}
\frac{2 n_N \sigma}{\sqrt{1 - (J/rw)^2}} \left(1 - 
\frac{\mu u^2}{4 w^2}\right)
\Theta\left(\frac{4}{\mu} - \frac{u^2}{w^2}\right)  ,
\end{equation}
where $w$ now stands for $u^2 + v^2$ where $v$ is the escape velocity
at radius $r$, and $\mu = m_X / m_N$ where $m_X$ is the mass of the
dark matter particle (in our case, the \simp) and $m_N$ is the average
nucleon mass. The first term in square brackets is a normalizing
factor, the term $2 n_N \sigma dr/\sqrt{1-(J/rw)^2}$ gives the
probability of collision, and the terms in the second line give the
probability that the particle gets scattered into a bound orbit.  The
step function $\Theta(x)$ equals $1$ or $0$ depending on whether its
argument is positive or not. It is simple to show that a collision
produces a fractional energy change $\Delta E/E$ that is uniformly
distributed between $0$ and $4/\mu$ where we have taken the large
$\mu$ limit (otherwise, $4/\mu$ is replaced by $4 \mu/ (\mu+1)^2$).

The above assumes that the dark matter particle typically suffers at
most one collision in its passage through the sun. We are here
interested in the opposite regime where $n_N \sigma R_\odot \gg 1$.
For each collision, the fractional energy change is of the order of
$1/\mu$. Therefore, after $n_N \sigma R_\odot$ collisions in the
sun, the \simp\ would get captured if $v_\odot^2 \, \simlt \,
(v_\odot^2 + u^2) (1 - 1/\mu)^{n_N \sigma R_\odot}$, implying
all particles with $u < u_*$ get captured where 
$u_* = v_\odot [1/(1 - 1/\mu)^{n_N \sigma R_\odot} - 1]^{1/2}$.
For capture by the sun, it adequate to approximate $u_*$
by $u_* = v_\odot / \sqrt {\mu/(n_N
\sigma R_\odot) - 1}$ if $\mu/{n_N \sigma R_\odot} > 1$ or $u_\star =
\infty$ if $\mu/{n_N \sigma R_\odot} \le 1$.
In other words, we
approximate $P$ in Eq.\ (\ref{C2}) as a step function.

\begin{figure}
\centering \leavevmode \epsfxsize=375pt \epsfbox{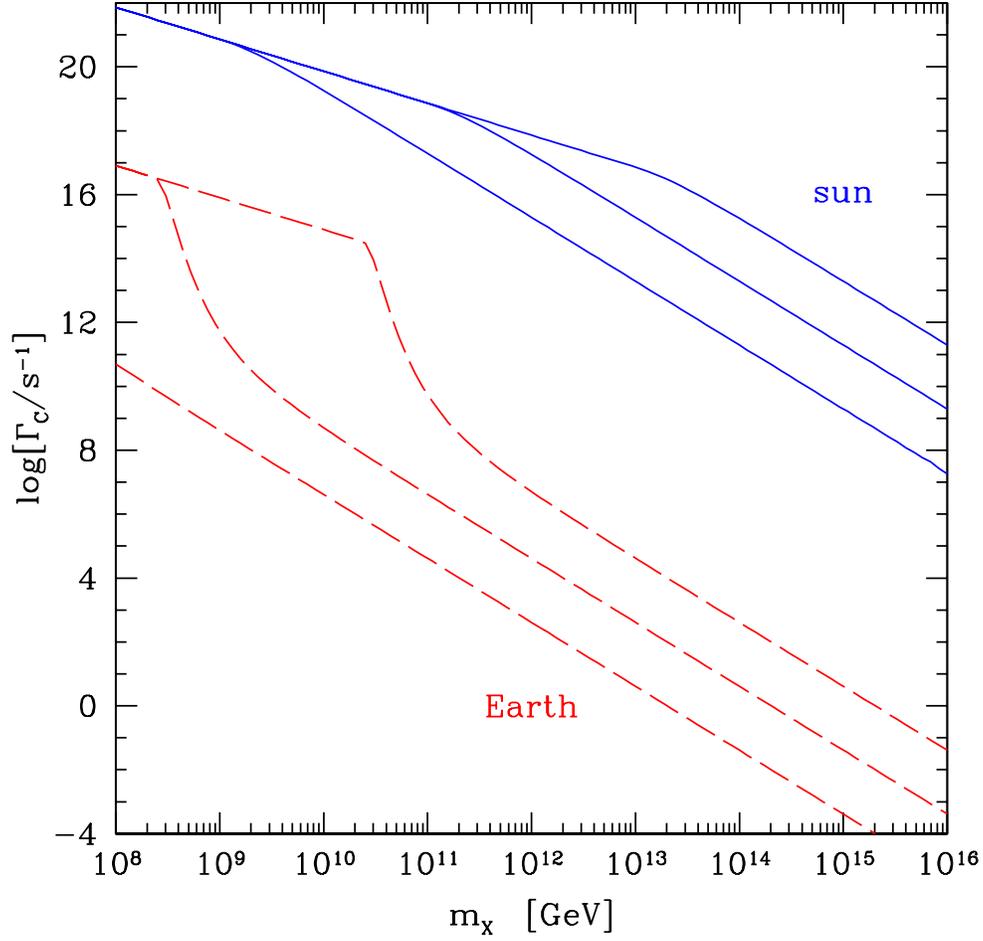}
\caption[fig1]{\label{fig:capture} The capture rate by the sun (solid curves) 
and Earth (dashed curves) for different values of the \simp--nucleon
cross section.  The three curves for the sun and for Earth correspond
to, from top to bottom, $\sigma = 10^{-22},\ 10^{-24},\ {\rm and}\
10^{-26}\
\cm^{2}$.}
\end{figure}

Using the Maxwell--Boltzmann phase space distribution function
for the \simps\,  $f(u) = 4 (n_{X}
/\sqrt{\pi u_{\rm th}^2})(u/u_{\rm th})^2 \exp(-u^2/u_{\rm th}^2)$
\cite{noteVel} with $\rho_{X} = 0.3(\GeV/m_X) \cm^{-3}$, 
we obtain from Eq.\ (\ref{C2}) the capture rate for \simps.  The
capture rate has two forms, depending on the efficiency of energy loss
in the sun.  This is parametrized by $q$, defined as
\begin{equation}
q \equiv \frac{\mu}{n_N\sigma R_\odot} = 20 
\left(\frac{m_X}{10^{12}\,\GeV} \right) 
\left(\frac{10^{-24}\,\cm^{2}}{\sigma}\right) 
\left(\frac{R_\odot}{7\times10^{10}\,\cm}\right)^2 
\left(\frac{2 \times 10^{33} {\,\rm g}}{M_\odot}\right).
\label{qdef}
\end{equation}
If $q\le 1$, the \simp\ will be efficient in losing energy in its passage
through the sun, and the capture rate is
\begin{equation}
\Gamma_C = 10^{17}\,\s^{-1} (1 + y^2) 
\left(\frac{10^{12}\,\GeV}{m_X}\right) 
\left(\frac{u_{\rm th}}{240\,\km\,\s^{-1}}\right) 
\left(\frac{R_\odot}{7\times10^{10}\,\cm}\right)^2 ,
\label{Cfinal2}
\end{equation}
where 
\begin{equation}
y \equiv 2.5 \left(\frac{v_\odot}{600\, \km\,\s^{-1}}\right)
\left(\frac{u_{\rm th}}{240\,\km\,\s^{-1}}\right)^{-1} .
\end{equation}
On the other hand, if $q>1$, only low velocity \simps\ are captured
and the capture rate is
\begin{equation}
\label{Cfinal1}
\Gamma_C = 10^{17}\,\s^{-1}\, 
\left[1 + y^2  - \exp(-x^2) (1 + y^2 + x^2) \right] 
\left(\frac{10^{12}\,\GeV}{m_X}\right)
\left(u_{\rm th} \over 240\,\km\,\s^{-1} \right) 
\left(\frac{R_\odot}{7\times10^{10}\,\cm}\right)^2
\end{equation}
where
\begin{equation}
x \equiv \frac{y}{\sqrt{q -1}} .
\end{equation}
The capture rate as a function of \simp\ mass and cross section is
shown in Fig.\ \ref{fig:capture}.

The captured \simps\ settle into the core of the sun on a time scale
determined by their drift velocity, $t_{\rm drift} \sim r / v_{\rm
drift}$ where $v_{\rm drift}$ can be estimated by balancing gravity
with the viscous drag due to scattering with nucleons \cite{gould90}:
$G \bar\rho r^3 m_X / r^2 \sim \sigma n_N v_{\rm drift} m_N (\bar{T}
/m_N)^{1/2}$. Here, $\bar{\rho}$ and $\bar{T}$ is the typical mass
density and temperature of the sun.  The timescale is very short, of
order $100\,\s\ (\sigma/10^{-24}\,\cm^2)(10^{12}\,\GeV/m_X)
(\bar{T}/10^7\,\K)^{1/2}$.

The subsequent evolution of the collection of \simps\ at the core can
be divided into two stages.  The first stage is when $N$, the total
number of \simps, is less than $N_{SG}$, the critical number necessary
for the \simps\ to become self gravitating.  In this first stage the
\simps\ are supported by the thermal pressure of the surrounding
plasma and the \simp\ profile follows an isothermal distribution given
by $\rho_X(r) \propto \exp(-r^2/2r_*^2)$ where $r_*= (3T_\odot/4 \pi G
m_X \rho_\odot)^{1/2}=5000\,\cm \ (10^{12}\,\GeV/m_X)^{1/2}
(T_\odot/10^7\,\K)^{1/2}(150\,\g\,\cm^{-3} /\rho_\odot)^{1/2}$, and
$T_\odot$ and $\rho_\odot$ are the core temperature and density
respectively. The critical number is defined by $4 \pi r^3_*
\rho_\odot /3 \sim N_{SG} m_X$, and is equal to
\begin{equation}
N_{SG} \sim 10^{26} 
\left(\frac{10^{12}\,\GeV}{m_X}\right)^{5/2} 
\left(\frac{T_\odot}{10^7\,\K}\right)^{3/2}
\left(\frac{150\,\g\,\cm^{-3}}{\rho_\odot}\right)^{1/2}  .
\label{Ng}
\end{equation}
To determine if $N$ ever exceeds $N_{SG}$, we have to determine
$N_{EQ}$, the number of \simps\ in the sun when there is an
equilibrium between annihilation and capture.  If the size is given by
$r_*$ above, then $N_{EQ}$ is found by equating the annihilation
rate and the capture rate: $2 \langle \sigma_A v\rangle (N_{EQ})^2 
/ (4\pi r_*^3/3) \sim \Gamma_C$.  Using the annihilation
cross-section $\langle \sigma_A v\rangle \simlt 1/(m_X^2 v)$, we obtain
\begin{equation}
N_{EQ} \sim 10^{30} 
\left(\frac{m_X}{10^{12}\,\GeV}\right)^{1/2}
\left(\frac{T_\odot}{10^7\,\K}\right)^{3/2}
\left(\frac{150\,\g\,\cm^{-3}}{\rho_\odot}\right)^{3/2}
\left(\frac{\Gamma_C}{10^{17}\,\s^{-1}}\right)^{1/2}
\left(\frac{v}{10^{-9}c}\right)^{1/2}     .
\label{Ne}
\end{equation}
The above means that unless $m_X$ is significantly smaller than
$10^{12}$GeV, no equilibrium is reached for $N < N_{SG}$.
With the capture rate given above, $N$ can reach
$N_{SG}$ on a time-scale much shorter than the lifetime
of the sun. The next stage is then set for the collapse of
the \simp\ collection, whose final state will be determined
by a number of factors.

First, the critical $N_{\rm Chandra}$, beyond which the collection of
\simps\ cannot be supported by degeneracy pressure, is given by
balancing $G m_X^2 N_{\rm Chandra} / r \sim \alpha^{1/3} N_{\rm
Chandra}^{1/3}/r$:
\begin{equation}
N_{\rm Chandra} \sim \alpha^{1/2} 10^{21} 
\left(\frac{m_X}{10^{12}\,\GeV}\right)^{-3}    ,
\label{chandraeq}
\end{equation}
where $\alpha$ is the ratio of electron number density to 
\simp\ number density. This is initially a large number, {\em i.e.},
the mass density of \simps\ and the mass density of nuclei are
initially comparable, which implies $\alpha \sim m_X / m_N$.  It is
unclear how much $\alpha$ would be reduced in the collapse process. It
depends on how effective \simps\ are in dragging along protons.  If
$\alpha \, \simgt \, 10^{10} (m_X/10^{12} {\, \rm GeV})$, $N_{\rm
grav}$ is smaller than $N_{\rm Chandra}$, and so the collapse would
result in a \simp\ collection supported by non-relativistic electron
degeneracy pressure. If $\alpha \simlt 10^{10} (m_X/10^{12}\,\GeV)$,
the configuration will collapse to a black hole, unless sufficient
annihilation occurs along the way.

Let us consider the question of whether annihilation would halt an
otherwise catastrophic collapse. Following
\cite{stark}, the fractional change in $N$ can be estimated by
\begin{equation}
{\Delta N \over N} \sim N 
\int dt \ 3 {\langle \sigma_A v \rangle \over 4 \pi r^3(t)}  .
\end{equation}
Whether sufficient annihilation occurs or not depends on whether the
integral is dominated by small $r$ (late times) or large $r$ (early
times). The longer the configuration spends at small radii, or in
other words, the slower the acceleration, the better the chance for
annihilation to work against collapse.  The relation $dr/dt \sim (r_0
/ r)^2 (r_0/t_{\rm drift})$ was used in Ref.\ \cite{stark}, where
$r_0$ is the initial radius and $t_{\rm drift}$ is the drift-time
given before (the timescale for collapse is set by viscous drag).  A
perhaps more reasonable limit to how fast a given shell can accelerate
is given by free fall: $dr/dt \propto 1/r^{1/2}$.  Integrating, one
obtains:
\begin{equation}
\frac{\Delta N}{N} \sim \frac{N}{2\pi} \frac{\langle \sigma_A v \rangle 
t_{\rm drift}}{r_0^{3/2} r_f ^{3/2}}   ,
\label{dNN}
\end{equation}
where we have taken the limit of $r_f \ll r_0$. Clearly, if $r_f$ were
sufficiently small, $\Delta N / N \sim 1$ can always be achieved. The
only thing one has to make sure is that the required $r_f$ is larger
than the Schwarzschild radius. Using $r_0 \sim r_*$ and $N \sim
N_{SG}$, it can be verified that
\begin{equation}
r_f \sim 10^{-4}\,\cm \left(\frac{N}{10^{30}}\right)^{2/3}
\left(\frac{\langle \sigma_A v \rangle}{10^{-32}\,\cm^{3}\,\s^{-1}}\right)
^{2/3}
\left(\frac{t_{\rm drift}}{100\,\s}\right)^{2/3}
\left(\frac{5000\,\cm}{r_0}\right)
\end{equation}
would do the job.

After shedding a fair fraction of the \simps, the equilibrium
configuration should in principle be one where support is provided by
non-relativistic electron degeneracy pressure: $G m_X^2 N /r \sim
\alpha^{2/3} N^{2/3} / (2 m_e r^2)$ where $m_e$ is the electron mass.
The above, together with the requirement of the balance of
annihilation and capture, gives
\begin{eqnarray}
\label{Nef}
N_{EQ} \sim 10^{20} \alpha^{2/3}
\left(\frac{\Gamma_C}{10^{17}\,\s^{-1}}\right)^{1/3}
\left(\frac{10^{12}\,\GeV}{m_X}\right)^{4/3}
\left(\frac{5 \times 10^{-4}\,\GeV}{m_e}\right)
\left(\frac{v}{10^{-9}c}\right)^{1/3}   .
\end{eqnarray}

Comparing the above with $N_{\rm Chandra}$ shows that the final
configuration is just barely stable, depending somewhat on the exact
value of $\alpha$ and $m_X$. If not stable, then the configuration
goes through another cycle of collapse and eventual halt by
annihilation. It is curious that this cycle might go on indefinitely,
in which case each collapse would be accompanied by enhanced
annihilation and therefore a mild neutrino outburst (using for example
$\Delta N \sim 10^{20}$ and $t_{\rm drift} \sim 100$ s gives an
annihilation rate of $10^{19}\,\s^{-1}$, not overwhelmingly larger
than the capture rate of $\sim 10^{17} \s^{-1}$). We will assume for the
rest of this paper that the annihilation rate is given by capture
rate, or more precisely, $2 \Gamma_A \sim \Gamma_C$.

Finally, there is the possibility that $\alpha$ drops to a
sufficiently small value later on that electron degeneracy pressure is
irrelevant, and the \simps\ are supported instead by their own
degeneracy pressure (or its analog if it were a boson; see {\em e.g.,}
Ref.\ \cite{liddle97}).  In this case the expression equivalent to 
Eq.\ (\ref{chandraeq}) is 
\begin{equation}
N_{\rm Chandra} \sim 10^{21}
(10^{12}\,\GeV/m_X)^{3},
\end{equation}
and the equilibrium number will be 
\begin{equation}
N_{EQ} \sim 10^5 (\Gamma_C /
10^{17}\,\s^{-1})^{1/3}(10^{12}\,\GeV/m_X)^{7/3} (v/10^{-9}c)^{1/3} ,
\end{equation} 
so that the final equilibrium configuration is clearly stable.

For completeness, we give the capture rate of \simps\ by Earth:
\begin{eqnarray}
\label{GammaCEarth}
\Gamma_C 
& = & 8\times10^{12}\,\s^{-1}
     \left[1 + y^2  -  \exp(-x^2) (1 + y^2 + x^2)\right]
     \left(\Frac{10^{12}\,\GeV}{m_X}\right) , 
\end{eqnarray}
where now $y=0.04$, and $x$ is given by
\begin{equation}
x \equiv y \left[{1\over (1- \mu^{-1})^{N_{\rm coll}}}-1 \right]^{1/2}, \ \ 
N_{\rm coll} \equiv 2 \times 10^{9} 
\left({\sigma \over 10^{-24} {\, \rm cm^2}}\right), \ \ 
\mu \equiv 10^{12} \left(m_X \over {10^{12} {\, \rm GeV}}\right).
\end{equation}
The parameter $N_{\rm coll}$ is the average number of collisions the
\simp\ suffers.  It may be identified with $\mu/q$ in the case of the sun
\cite{rock}.  The expression for $\Gamma_C$ is well approximated by the
corresponding expressions in Eqs.\ (\ref{qdef}) through
(\ref{Cfinal1}) for $y > 1$, which applies for the case of the
sun. For Earth we have to resort to this more complicated expression;
however, it has simple limiting forms.  For $N_{\rm coll}/\mu \gg 1$,
one can use [cf.\ Eq.\ (\ref{Cfinal2})]
\begin{equation}
\Gamma_C 
= 8\times10^{12}\,\s^{-1} (1 + y^2) \left(\Frac{10^{12}\,\GeV}{m_X}\right) .
\end{equation}
For $N_{\rm coll}/\mu \ll 1$, the capture rate is well approximated by
[cf.\ Eq.\ (\ref{Cfinal1})]
\begin{equation}
\Gamma_C = 8\times10^{12}\,\s^{-1}\  x^2 y^2 
\left(\Frac{10^{12}\,\GeV}{m_X}\right)
\end{equation}
with $x = y/\sqrt{q - 1}$, where $q = 5 \times 10^2 (m_X/10^{12}\GeV)
(10^{-24}\cm^{2}/\sigma)$.  For most, but not all, of the parameters
of interest, it is this last regime that is relevant for capture by
Earth, which gives $\Gamma_C \sim 4 \times 10^{4}\s^{-1}
(10^{12}\GeV/m_X)^2 (\sigma/ 10^{-24}\cm^{2})$.  The total number of
\simps\ captured in the lifetime of Earth, $t_E \sim 10^{17}\s$, 
is then $4 \times 10^{21} (10^{12}\GeV/m_X)^2 (\sigma/10^{-24}\cm^{2})$.
We can compare this with the number of \simps\ in equilibrium using Eq.\ 
(\ref{Ne}), with the replacements $T_\odot \rightarrow T_\oplus
\simeq 5000$ K, $\rho_\odot \rightarrow \rho_\oplus \simeq 10\ \g\ \cm^{-3}$,
and $v \rightarrow 2 \times 10^{-11}c$, obtaining $N_{EQ} \sim 5\times 10^{18}
(m_X / \GeV)^{1/2}(\Gamma_C/ 400\ \s^{-1})^{1/2}$.  Therefore, the \simps\ 
captured in Earth would be in equilibrium unless the mass or cross section 
is significantly different from what is assumed above.  In general, the
annihilation rate is given by:
\begin{eqnarray}
\Gamma_A = 2 \times 10^4 {\, \rm s^{-1}} \left[\frac{\Gamma_C} 
{4 \times 10^4\s^{-1}}\right] \tanh^2 
\left[10^3 \left(\frac{10^{12}\GeV}{m_X}\right)^{1/2}
\left(\frac{\Gamma_C}{4 \times 10^4\s^{-1}} \right)^{1/2}
\left(\frac{t_E}{10^{17}\s}\right) \right] .
\end{eqnarray}

The capture rate for Earth is shown in Fig.\ \ref{fig:capture} along
with the capture rate for the sun.

\section{SIMPZILLA ANNIHILATION IN THE SUN}

Take a simple picture where the \simp\ annihilates and produces two
quarks or two gluons, each of energy $m_X$ where $m_X$ is the mass of
the \simp.  The quarks and gluons then fragment into high
multiplicity jets of hadrons and secondary decay products.

Defining $x=E/E_{\rm jet} = E/m_X$, one can take the fragmentation
function for the total number of hadrons to be \cite{chill}
\begin{equation}
\frac{dN_H}{dx} = a x^{-3/2}(1-x)^2.
\end{equation}
Here $a$ is some constant that can be set by total energy
considerations: 
\begin{equation}
1=\int_0^1 x \frac{dN_H}{dx} dx = a \frac{16}{15},
\end{equation}
so $a = 15/16$.

The total number of hadrons produced in the fragmentation of the jet is
\begin{equation}
N_H = \int_\epsilon^1 \frac{dN_H}{dx}
dx=\frac{15}{8}\frac{1}{\sqrt{\epsilon}}
\left[ 1 - \frac{8}{3}\sqrt{\epsilon} + 2\epsilon -\epsilon^2/3\right].
\end{equation}
When calculating the total multiplicity, the cutoff for the integral
should be $\epsilon=\Lambda_{\rm QCD}/m_X$.  Using $\Lambda_{\rm
QCD}=0.1$\,GeV and defining $M_{12}=m_X/10^{12}$\,GeV,
\begin{equation}
\label{eq:ntotal}
N_H = \frac{15}{8}\left(\frac{m_X}{\Lambda_{\rm QCD}}\right)^{1/2} 
= 6\times10^6 M_{12}^{1/2} . 
\end{equation} 
The final decay chain of the hadrons will contain all species of neutrinos.

We will be interested in the number of heavy quarks, bottom and top.
If all quarks were light, then all flavors would be produced equally.
However, because of the mass of the heavy quark, the number must be
found by using $\epsilon=M_Q/m_X$ in Eq.\ (\ref{eq:ntotal}) rather
than $\epsilon=\Lambda_{\rm QCD}/m_X$:
\begin{equation} 
N_Q \simeq N_H \sqrt{\frac{\Lambda_{\rm QCD}}{M_Q}}, 
\end{equation}
where $M_Q$ is the mass of a typical meson containing the heavy quark:
2 GeV for the charm, 5 GeV for bottom, and 175 GeV for top.
Therefore, per annihilation, one expects $N_{\rm charm}/N_H=0.23$,
$N_{\rm bottom}/N_H=0.14$, and $N_{\rm top}/N_H=0.024$.  This means
that each annihilation into two jets produces $7.4\times10^6
M_{12}^{1/2}$ light hadrons, $2.8\times10^6 M_{12}^{1/2}$ charmed
hadrons, $1.6\times10^6 M_{12}^{1/2}$ bottom hadrons, and
$2.8\times10^5 M_{12}^{1/2}$ top hadrons.

It is also possible to estimate the spectrum of the heavy-quark
hadrons: 
\begin{equation}
\frac{E_{\rm min}}{N_{\rm TOTAL}} \ \frac{dN}{dE} \sim 
\frac{1}{2} \left( \frac{E}{E_{\rm min}} \right)^{-3/2}, 
\qquad (E> E_{\rm min})  ,
\end{equation}
where $E_{\rm min}$ is the mass of the top quark, bottom quark, or charm
quark. 

\begin{figure}
\centering \leavevmode \epsfxsize=450pt \epsfbox{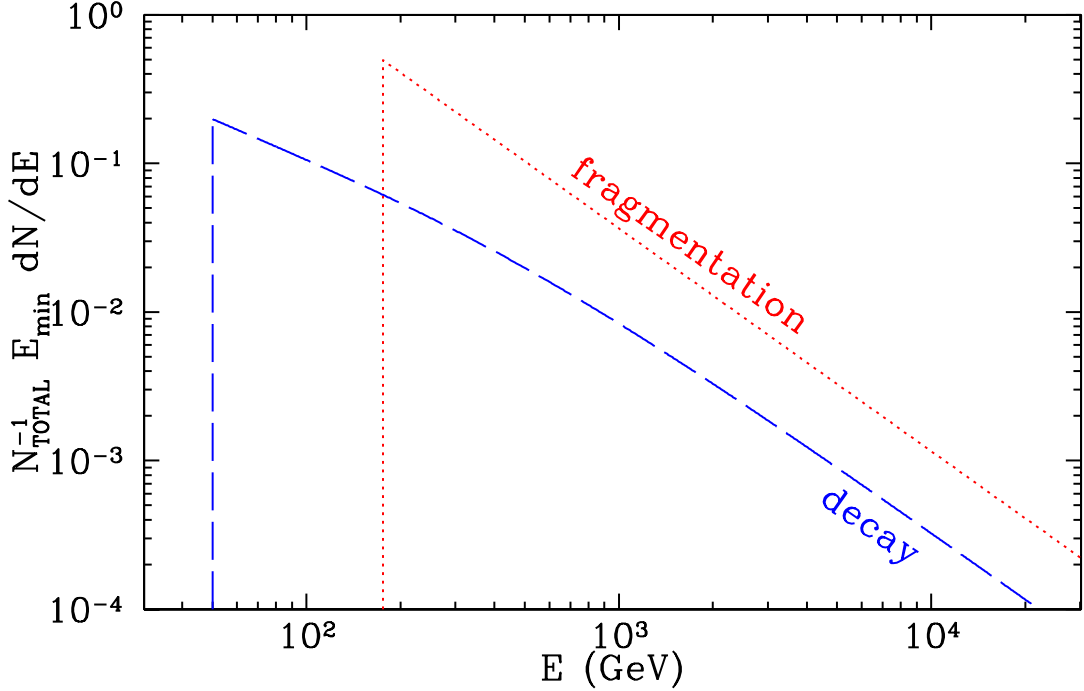}
\caption[fig2]{\label{fig:spectrum} The spectrum of top hadrons produced
by the fragmentation of quark and gluon jets from \simp\ annihilation
(dotted line), and the \nt\ spectrum produced by decay of top (dashed
line).  For the fragmentation spectrum, $E_{\rm min}=175$ GeV and
$N_{\rm TOTAL}=2.8\times10^5M_{12}^{1/2}$ per annihilation.  For the
decay spectrum, we will only be concerned with neutrinos above $E_{\rm
min}=50$ GeV.  The number of tau neutrinos above $E_{\rm min}$ is
$N_{\rm TOTAL}=10^4M_{12}^{1/2}$ per annihilation, and half that for
the other neutrinos.}
\end{figure}

The resulting $E^{-3/2}$ fragmentation spectrum for top hadrons is
shown by the dotted line labeled ``fragmentation'' in Fig.\
\ref{fig:spectrum}. The minimum energy is approximately the mass of
the top quark, 175 GeV.

We will assume that \simp\ annihilation occurs in a medium of
density found in the center of the sun, $\rho \sim 200\,\g\,\cm^{-3}$,
or $n \sim 10^{26}\,\cm^{-3}$.  Using an interaction cross section of
$10^{-24}\,\cm^2$, the hadronic interaction length is about
$10^{-2}\,\cm$.  Light and charmed hadrons scatter many times
before decay and the resultant neutrinos will have very low
energy. The $B$ lifetime is about $10^{-12}\,\s$, so the decay
length is $3\times10^{-2}(E/M_B)\,\cm$, or
$6\times10^{-3}(E/\GeV)\,\cm$ using $M_B=5\,\GeV$.  The ratio of the
decay length to interaction length is $L_D/L_I=0.6(E/\GeV)$.  So for
$E>5\,\GeV$ the $B$ will also scatter and lose energy before decay.
It won't completely stop after one scattering.  Most of the cross
section is diffractive production of low-energy crud and there should
still be a leading $B$, so $B$ decay is a potential source of
high-energy neutrinos.

However, top hadrons are a promising source.  The top lifetime is
short, and almost 100\% of the time decays as $t \rightarrow Wb$.  The
$W$ lifetime is $3\times10^{-25}(E/80\GeV)$ s, which results in a
decay length of about $10^{-16}(E/\GeV)\cm$, which will be much less
than the interaction length for even the most energetic tops.  The $W$
then decays with a branching ratio of $1/3$ as $W\rightarrow l\nu_l$
(equally into $\tau\nu_\tau$, $\mu\nu_\mu$ and $e\nu_e$).  Therefore
it is reasonable to assume the top quark will produce energetic
neutrinos before losing energy.

The spectrum of neutrinos produced in the chain $t\rightarrow W
\rightarrow \nu$ is straightforward to calculate.  A convenient analytic 
fit to the spectrum is given by
\begin{equation}
\label{eq:labor}
\frac{dN}{dE} = {\cal N} \ \frac{E+M_W}
{\sqrt{ \left[E+M_t\right] \left[(E+M_t)^2-M_t^2\right]
\left[(E+M_W)^2-M_W^2\right] } } ,
\end{equation}
where ${\cal N}$ is a normalization factor. This spectrum is shown in
Fig.\ \ref{fig:spectrum}.  As expected, at energies larger than the top
mass the $E^{-3/2}$ spectrum is recovered.

We will use the spectrum in Fig.\ \ref{fig:spectrum} as the spectrum
of neutrinos produced by \simp\ annihilation.  Since there are about
$2.8\times10^5M_{12}^{1/2}$ tops produced per annihilation, and 10\%
of them make $\tau\nu_\tau$ followed by $\tau$ decay including a
$\nu_\tau$, the total yield of $\nu_\tau$'s per annihilation is
$5.6\times10^{4}M_{12}^{1/2}$. Top (as well as $\tau$) decay also
produces $\mu+\nu_\mu$ and $e+\nu_e$, but the electrons are absorbed
and the muons are stopped before decay, so the yield of high-energy
$\nu_\mu$ and $\nu_e$ per annihilation is about $3.8\times10^4
M_{12}^{1/2}$. Only about 20\% of the neutrinos will be produced with
energy above 50 GeV, so the emission rate of tau neutrinos in the core
is about $10^4 M_{12}^{1/2}\Gamma_A$ with a spectrum above 50 GeV
shown in the figure.  This, of course, is the emission rate in the
core.  We now turn to the propagation of the neutrinos through the
sun.

\section{The emergent spectrum of neutrinos}

The total neutrino emission rate above 50 GeV from the core of the sun
is $f_{\rm core} = 10^{4}M_{12}^{1/2}\Gamma_A$ for \nt, and half that
for $\nu_\mu$ and $\nu_e$. Here $\Gamma_A=\Gamma_C/2$ is the \simp\
annihilation rate.  The core emission rate spectrum above 50 GeV 
for each neutrino species is [see Eq.\ (\ref{eq:labor})]
\begin{eqnarray}
\left(\frac{df}{dE}\right)_{\rm core} & = & 333\ \GeV^{3/2} \
\frac{f_{\rm core}}{E_{\rm min}} \nonumber \\
& & \times \frac{E+M_W}
{\sqrt{ \left[E+M_t\right] \left[(E+M_t)^2-M_t^2\right]
                           \left[(E+M_W)^2-M_W^2\right] } }
	\ \Theta(E-E_{\rm min}),
\end{eqnarray}
where $E_{\rm min}=50$ GeV and the theta-function vanishes for
negative argument and is unity for positive argument.  Of course
$df/dE$ has been normalized such that
\begin{equation}
f_{\rm core} = \int_{E_{\rm min}}^\infty  
	\left(\frac{df}{dE}\right)_{\rm core} dE .
\end{equation}

But since the sun is opaque to energetic neutrinos, the emergent
emission rate spectrum is not the same as the core emission rate
spectrum.  The emission rate spectrum of neutrinos that emerge
unscathed is
\begin{equation}
\label{unscattered}
\left(\frac{df}{dE}\right)_{\rm unscattered} = 
\left(\frac{df}{dE}\right)_{\rm core}
\exp \left(-\sigma(E)\int_0^\infty n(r)\ dr \right) ,
\end{equation}
where $n(r)$ is the radial dependence of the number density of the sun
and $\sigma_{CC}(E)$ is the energy-dependent charged-current cross
section given in Table \ref{sigmas}.  Also given in Table \ref{sigmas}
is the neutral-current cross section.

\begin{table}
\caption[t1]{\label{sigmas} The energy-dependent cross sections 
used to calculate the flux of neutrinos from the sun \cite{gandhi}.}
\begin{ruledtabular}
\begin{tabular}{lcc}
Energy Range \ [GeV] & $\sigma_{NC}$ \ $[\cm^2]$ & $\sigma_{CC}$ \
$[\cm^2]$ \\
\hline \colrule
$ \phantom{1^4} 0 \le E \le 10^{4} $       
	& $ 2.0 \times 10^{-39} \left(\Frac{E}{\GeV}\right)\phantom{^{0.123}} $
	& $ 6.6 \times 10^{-39} \left(\Frac{E}{\GeV}\right)\phantom{^{0.123}} $
	 \\
$ 10^{4} \le E \le 10^{5} $
	& $ 2.1 \times 10^{-38} \left(\Frac{E}{\GeV}\right)^{0.714} $
	& $ 6.1 \times 10^{-38} \left(\Frac{E}{\GeV}\right)^{0.714} $
	 \\
$ 10^{5} \le E \le 10^{7} $
	& $ 0.3 \times10^{-36} \left(\Frac{E}{\GeV}\right)^{0.462}  $
	& $1.0 \times 10^{-36} \left(\Frac{E}{\GeV}\right)^{0.462}  $
	 \\
$ 10^{7} \le E \le 10^{12} $
	& $ 2.3 \times 10^{-36} \left(\Frac{E}{\GeV}\right)^{0.363} $
	& $ 5.5 \times 10^{-36} \left(\Frac{E}{\GeV}\right)^{0.363} $
\end{tabular}
\end{ruledtabular}
\end{table}

\begin{figure}
\centering \leavevmode \epsfxsize=450pt \epsfbox{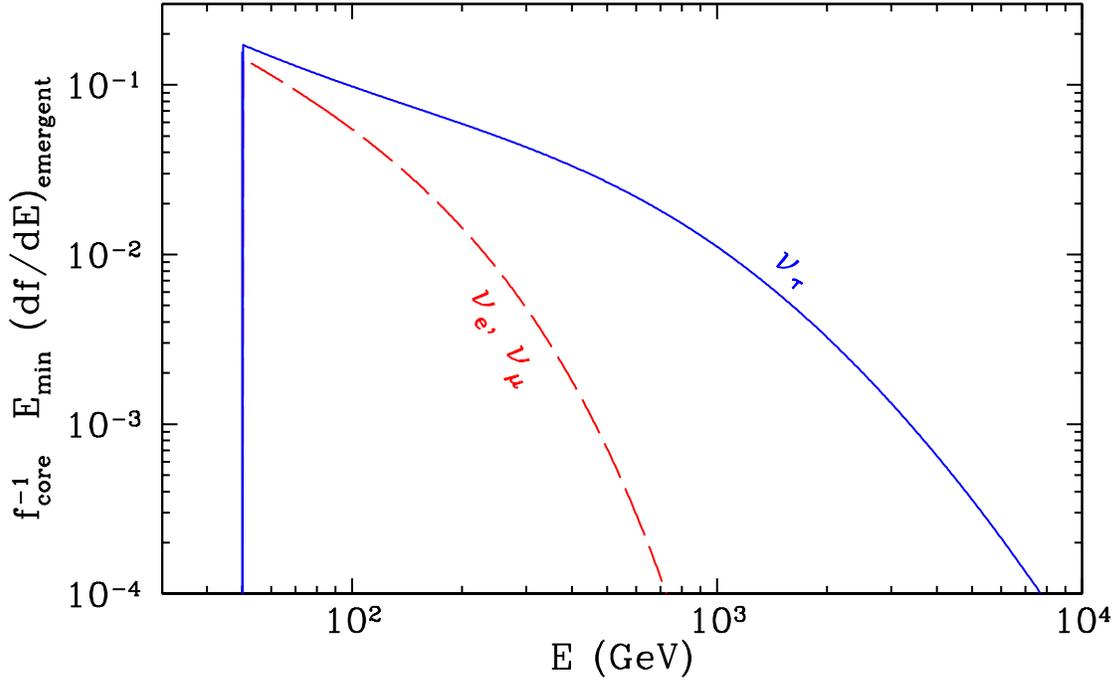}
\caption[fig3]{\label{fig:fluence} The emergent emission rate spectrum of 
neutrinos from the sun. For \nt, $f_{\rm core} = 
10^4M_{12}^{1/2}\Gamma_A$, and for $\nu_\mu$ and $\nu_e$, half that
value.  In all cases, $E_{\rm min} = 50$ GeV. }
\end{figure}

The electrons and muons produced by charged-current interactions are
rapidly thermalized, so the effect of charged-current interactions is
effectively to remove $\nu_\mu$ and $\nu_e$ neutrinos above a
transparency energy $E_\kappa$ where $\sigma(E_\kappa)\int n(r)\ dr$
becomes unity. Using
\begin{equation}
n(r) = 1.4\times10^{26}\exp(-r/0.1R_\odot)\ \cm^{-3}
\end{equation}
for the density profile and adopting the cross section for $E<10^4$
GeV, the transparency energy is $E_\kappa=150$ GeV, and the emission
spectrum of $\nu_\mu$ and $\nu_e$ emergent from the sun is
\begin{equation}
\label{mueemissionspectrum}
\left(\frac{df}{dE}\right)_{\rm emergent} = 
\left(\frac{df}{dE}\right)_{\rm core} \exp(-E/E_\kappa) \ .
\end{equation}
This emission rate spectrum is shown in Fig.\ \ref{fig:fluence}.

The situation is different for tau neutrinos.  The lifetime of the tau
produced in a charged-current scattering is so short that it decays
before significant energy loss.  Since the decay of the tau includes a
\nt, the effect of the scattering and subsequent decay is simply to
reduce the energy of the incident \nt\ to about 20\% of its incident
value.  Therefore, an incident tau neutrino above the transparency
energy will continually suffer tau production and decay interactions,
but will not be removed from the flux of high-energy neutrinos.  The
process will continue until the tau-neutrino energy has been degraded
to the transparency energy or below, then the tau neutrino will
escape.

This process has been considered by Halzen and Saltzberg
\cite{francis} for very energetic neutrinos propagating through Earth.  
They found that the emergent spectrum of neutrinos was very well
described by a lognormal distribution centered on the transparency
energy with a dispersion of 0.49 decades in energy.  We will assume
that the flux of scattered tau neutrinos above the transparency energy
emerges in a lognormal distribution peaked at $E_\kappa=150\ \GeV$ with
a dispersion of 0.49 decades in energy.

The emergent emission spectrum of tau neutrinos has two contributions.
The first contribution is the unscattered emission spectrum, given by Eq.\
(\ref{unscattered}).  The second contribution is the fraction of the
original emission above the transparency energy which will emerge as a
lognormal distribution centered on the transparency energy;
\begin{equation}
\label{lognormal}
\left(\frac{df}{dE}\right)_{\rm scattered} = \frac{f_{\rm core}}{E_{\rm min}}\ 
\ {\cal F} \ 
\exp\left[-\left( \log E -\log E_\kappa\right)^2 / 2\sigma^2\right] ,
\end{equation}
where $\sigma = 0.49$ and ${\cal F}$ is found by demanding that the
integral of Eq.\ (\ref{lognormal}) results in the total number of
neutrinos above 150 GeV that are scattered. (If the initial energy of
the neutrino is below 150 GeV, a scattering will produce a neutrino
below $E_{\rm min}$.)  The result is ${\cal F}=4.6\times 10^{-2}$.

So for tau neutrinos, the emergent neutrino spectrum is 
\begin{equation}
\left(\frac{df}{dE}\right)_{\rm emergent} =
\left(\frac{df}{dE}\right)_{\rm unscattered} + 
\left(\frac{df}{dE}\right)_{\rm scattered} .
\label{tauemissionspectrum}
\end{equation}
This spectrum is also shown in Fig.\ \ref{fig:fluence}.

\section{The Event Rate}

In the last section we calculated the emission rate and emission rate
spectrum of neutrinos from the sun.  In this section we will calculate
the event rate in a suppositious underice or underwater neutrino
detector of approximate size of a cubic kilometer \cite{icecube}.  We
will only consider the rate for ``contained events,'' where the
neutrino converts inside the volume of the detector. Including
``uncontained events'' will not significantly alter our results
because the muon range at the relevant energy range here is comparable
to a kilometer.  We will assume that the efficiency of detection is a
step function: zero below 50 GeV and unity above 50 GeV.

The first step in calculating the event rate is the simple step of
converting the emission rate spectrum calculated in the last section
to a flux spectrum arriving at Earth.  Since the sun-Earth distance is
$D=1.5\times10^8$ km, the flux spectrum is
\begin{equation}
\frac{dF}{dE} = \frac{1}{4\pi D^2}\left(\frac{df}{dE}\right)_{\rm emergent}
=3.5\times10^{-18}\left(\frac{df}{dE}\right)_{\rm emergent} \ \km^{-2} .
\end{equation}

\begin{figure}
\centering \leavevmode \epsfxsize=450pt \epsfbox{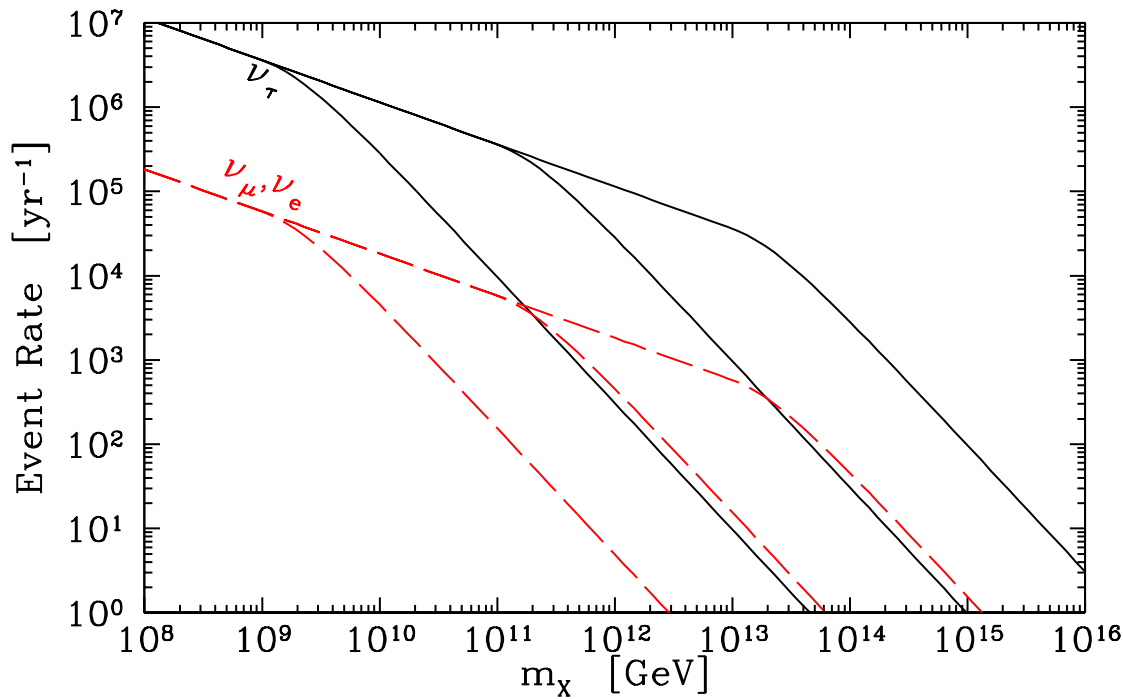}
\caption[fig4]{\label{fig:eventrate} The event rate in a
cubic-kilometer underice/underwater detector for different values of
the \simp--nucleon cross section: from top to bottom the curves are
for $\sigma = 10^{-22},\ 10^{-24},\ {\rm and}\ 10^{-26}\ \cm^{2}$. The
upper solid curves are for tau neutrinos and the lower dashed curves
are for muon and electron neutrinos. For comparison, the 90\% C.L.\
upper limits on muon fluxes of nonatmospheric origin in the direction
of Earth's core or the sun is approximately $10^{4}\ \km^{-2}\
\yr^{-1}$ \cite{boliev}.}
\end{figure} 

The mean-free-path of neutrinos is much larger than the size of the
detector, so the fraction of the incident neutrinos of energy $E$ that
convert inside the detector is $n_{\rm ice}\ \sigma_{CC}(E)\ L$ where
$n_{\rm ice}$ is the number density of the target and $L=1\ \km$ is
the size of the detector.  The event-rate spectrum is given simply by
\begin{equation}
\frac{dR}{dE} = \frac{dF}{dE}\left[n_{\rm ice}\ \sigma_{CC}(E)\ L \right] A
\ \Theta(E-50\ \GeV) ,
\end{equation}
where $A$ is the area of the detector, assumed to be 1 km$^2$, and the
$\Theta$-function represents the detector efficiency.

Since $n_{\rm ice}\ \sigma_{CC}(E)\ L=4\times10^{-10}(E/1\ \GeV)$, 
we find for the event-rate spectrum
\begin{equation}
\frac{dR}{dE} = 1.4\times10^{-27}\ \frac{E}{\GeV}\ 
\left(\frac{df}{dE}\right)_{\rm emergent}\ \Theta(E-50\ \GeV) .
\end{equation}
The total event rate is the integral of the event-rate spectrum.

\begin{figure} 
\centering \leavevmode \epsfxsize=450pt \epsfbox{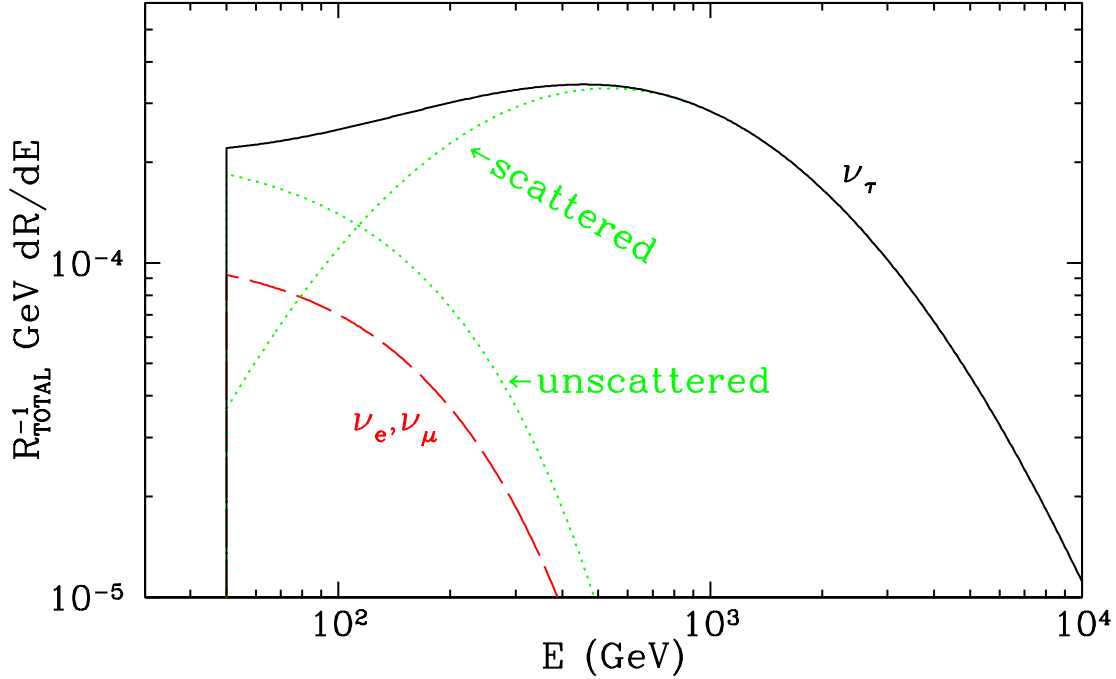}
\caption[fig5]{\label{fig:eventratespectrum} The.spectrum of events 
for different neutrino species.  The total event rate $R_{\rm TOTAL}
=1.1\times10^{-20}M_{12}^{1/2}\Gamma_A$ is the sum of the
electron-neutrino, muon-neutrino, and tau-neutrino rates. Also shown
are the two contributions to the tau-neutrino events, from unscattered
neutrinos and from scattered neutrinos [see Eq.\
(\ref{tauemissionspectrum})].}
\end{figure}

For muon or electron neutrinos, the total event rate can be found
using Eq.\ (\ref{mueemissionspectrum}) for the emergent emission
spectrum, with result
\begin{equation}
R_{\rm TOTAL}(\nu_\mu,\nu_e) = 1.6\times 10^{-22}M_{12}^{1/2}\Gamma_A .
\end{equation}
This event rate is shown in Fig.\ \ref{fig:eventrate}, and the
event-rate spectrum is shown in Fig.\ \ref{fig:eventratespectrum}.

For tau neutrinos, the total event rate is found using Eq.\
(\ref{tauemissionspectrum}) for the emergent emission spectrum.  
For tau neutrinos the total event rate is 
\begin{equation}
R_{\rm TOTAL}(\nu_\tau) = 1.1\times10^{-20}M_{12}^{1/2}\Gamma_A .
\end{equation}
This event rate is also shown in Fig.\ \ref{fig:eventrate}, along with
the event-rate spectrum in Fig.\ \ref{fig:eventratespectrum}.

The contribution of scattered tau neutrinos dominates the event rate.
While the $\nu_\mu$ and $\nu_e$ events would be peaked toward the lower
energy of the detector and drop rapidly, the $\nu_\tau$ events would
have a relatively flat spectrum extending from the lower limit of the
detector out to about 1000 GeV.

The mean energies of the detected neutrinos are
\begin{eqnarray}
\langle E \rangle = \phantom{3}197\ && \GeV\ \quad (\nu_\mu,\nu_e) \nonumber \\
\langle E \rangle = 3454\ && \GeV\ \quad (\nu_\tau) .
\end{eqnarray}

\begin{figure}
\centering \leavevmode \epsfxsize=450pt \epsfbox{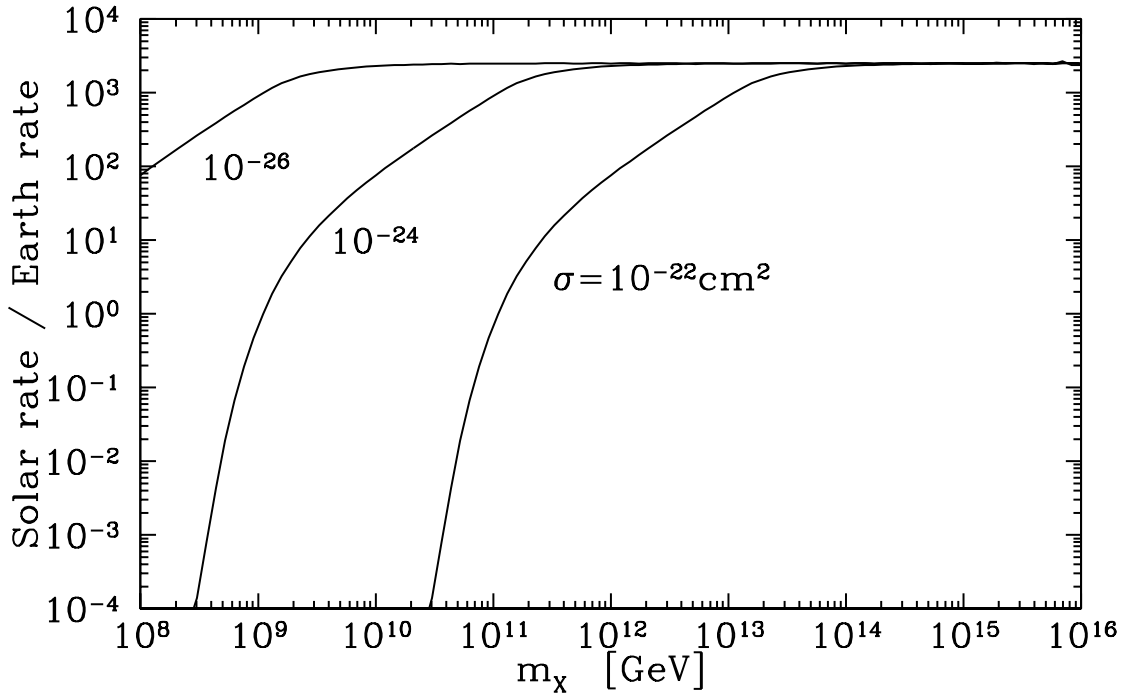}
\caption[fig6]{\label{fig:sunearth} The ratio of the event rates from 
solar and terrestrial neutrinos originating from \simp\
annihilation. The ratio is a function of \simp\ mass $m_X$ and
scattering cross section $\sigma$ through the dependence of the
capture rates on $m_X$ and $\sigma$. }
\end{figure}

We close this section by remarking on the possibility of a detectable
event rate from annihilation of \simps\ captured by Earth.  The ratio
of the event rate for neutrinos of solar origin to the event rate for
neutrinos of terrestrial origin is shown in Fig.\ \ref{fig:sunearth}.
For small mass or large interaction cross section the signal from the
center of Earth my be larger than the solar signal.  For $\sigma
\simlt 10^{-24}\ \cm^2$ the solar signal will dominate for \simp\
masses larger than about $10^9$ GeV, while for $\sigma \simlt 10^{-26}\
\cm^2$, the solar signal dominates for the entire range of \simp\ mass
considered here.

\section{Conclusions}

If the local dark matter is very massive and strongly interacting, it
should collect in the sun and Earth in sufficient numbers that
equilibrium will be maintained between the capture rate and the
annihilation rate.

Annihilation or decay of very massive particles into hadronic channels
in the solar core or at the center of Earth leads to the production of
high-energy neutrinos.  While electron and muon neutrinos above 150
GeV are mostly absorbed, energetic tau neutrinos will be emitted, and
will be the signature of energetic jet fragmentation at the center of the
sun.

For most of the range of parameter space, the event rate expected in
kilometer-scale neutrino detectors will be well within detection
limits.  Such detectors should be able to exclude (or confirm!) the
possibility of \simps\ as dark matter.

\begin{figure} 
\centering \leavevmode \epsfxsize=450pt \epsfbox{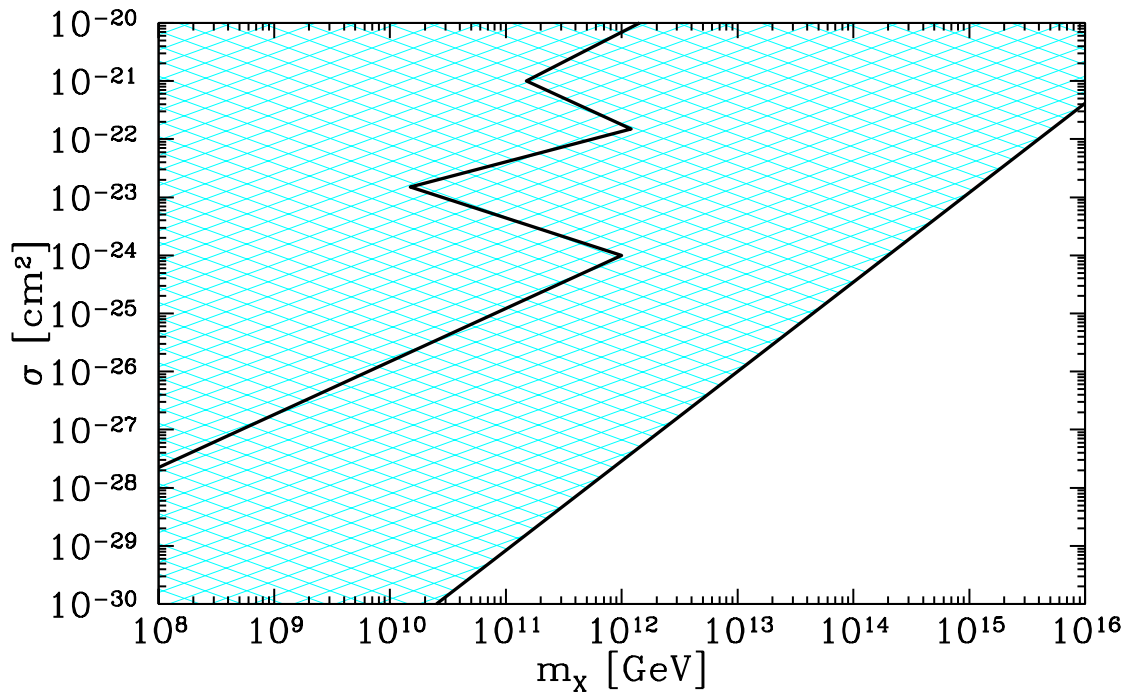}
\caption[fig7]{\label{fig:exclude} The shaded region above the jagged line is
excluded by a variety of considerations as discussed in \cite{stark}.
Slightly stronger, model-dependent constraints can be found in Refs.\
\cite{stark,bprs}.  The shaded region above the straight line would result 
in more than 10 events per year in a cubic-kilometer underice or
underwater detector and should easily be able to be excluded.}
\end{figure} 

In this paper we have only considered production of neutrinos through
top quark production and decay.  Another potential source of tau
neutrinos is bottom quark production and decay.  While this may
increase the emission rate of tau neutrinos, the spectrum is expected
to be the same.

It is important to note that there is essentially no background. For
example consider the background from cosmic-ray produced neutrinos.
To produce a 1 TeV neutrino, a center-of-momentum energy of around
$\sqrt{s}=\sqrt{2m_{\rm proton}E}=10$ TeV is required, where $E$ is
the cosmic ray energy.  Thus, a threshold energy of $E_{TH}=5\times
10^{16}$eV is required to produce a 1 TeV neutrino.  The cosmic-ray
flux at high energies is approximately
\begin{equation}
E^3 \ \frac{dF}{dE}=10^{24.5}\ \frac{{\rm eV^2}}{\s\ {\rm sr\ m}^2},
\end{equation}
which when integrated to give the total flux of particles with energy
greater than $E$ gives
\begin{equation}
F(E>E_{TH}) = \frac{1}{2E_{TH}^2}\ 10^{24.5}\ 
\frac{{\rm eV}^2}{\rm{s\ sr\ m^2}} = 
\frac{6.1}{{\rm yr\ deg}^2\ \km^2} .
\end{equation}
Since the sun subtends about a square degree, this sets the scale of
the background.  For much of parameter space, the signal should be
well above the background.

In Fig.\ \ref{fig:exclude} we present out results in the $\sigma$ vs.\
$m_X$ plane and compare then with other limits.  Clearly our
considerations greatly extends the excluded region.

Indirect detection of \simps\ through annihilation in the sun is
complementary to the other idea for indirect detection: {\sc
wimpzilla} decay producing ultra-high energy cosmic rays
\cite{first,second}.

Finally, we comment on the possible role of neutrino oscillations.
Neutrino oscillations will be important if
\begin{equation}
\frac{\Delta m^2}{10^{-3}{\rm eV}^2} \ \frac{10^2\GeV}{E_\nu} \ 
\frac{L}{2\times10^9\cm} \simgt 1,
\end{equation}
where $L$ is the path length. Since $2\times10^9\cm$ is $R_\odot/35$,
if $\Delta m^2$ is greater than or of the order of $3\times
10^{-5}{\rm eV}^2$, oscillations will occur in the sun.  If $\Delta
m^2$ is greater than or of order of $10^{-7}{\rm eV}^2$, then
oscillations will be important during the neutrino's transit to Earth.
Oscillations between $\nu_\tau$ and $\nu_\mu$ or $\nu_e$ in the sun
will decrease the $\nu_\tau$ emission rate, while oscillations in
transit will change the flavor signature of the signal.

\begin{acknowledgements}
I.F.M.A. was supported in part by NSF KDI Grant 9872979. The work of
E.W.K. was supported in part by NASA (grant number NAG5-7092).
L.H. was supported by NASA NAG5-7047, NSF PHY-9513835, and the Taplin
Fellowship at the IAS, and by the Outstanding Junior Investigator
Award, DOE grant DE-FG02-92-ER40699.  We would like to acknowledge
useful conversations with Pasquale Blasi, Gustavo Burdman, Francis
Halzen, Chris Hill and George Smoot. I.F.M.A. would like to
acknowledge the hospitality of Boston University.
\end{acknowledgements}



\bibliography{apssamp} 

\begin{thebibliography}{99}
\bibitem{crosssection} It is likely that the \simp\ will pick up a nucleus, 
so the cross section will be closer to a nuclear cross section rather
than a nucleon cross section.
\bibitem{noteVel} We ignore the velocity of the sun with respect
to the halo. This typically introduces a small suppression of the
capture rate (see \cite{gould87}).
\bibitem{rock} In the calculation of the energy loss through the 
sun and Earth we should include the effect of scattering from nuclei.
While this is not an important correction for the sun, it may make a
noticeable change in the energy loss through Earth.  The $A$
dependence of the cross section will have only a small effect, but the
mean energy loss per collision will scale with $A$..  This would have
the effect of decreasing $\mu$ by a factor of $\langle A \rangle$.
\end{thebibliography}

\end{document}